\documentclass[conference]{IEEEtran}
\IEEEoverridecommandlockouts
\usepackage{amsmath,amssymb,amsfonts}
\usepackage{algorithm}

\usepackage{subcaption}

\usepackage{graphicx}
\usepackage[hidelinks]{hyperref}
\usepackage{float}
\usepackage{svg}
\usepackage{circuitikz}
\usepackage{tikz}
\usetikzlibrary{shapes.geometric, arrows}
\usepackage{booktabs}
\usepackage{multirow}
\usepackage{algpseudocode}
\algnewcommand{\LeftComment}[1]{\Statex \(\triangleright\) #1}
\usepackage{textcomp}
\usepackage{xcolor}
\usepackage[english]{babel}
\def\BibTeX{{\rm B\kern-.05em{\sc i\kern-.025em b}\kern-.08em
    T\kern-.1667em\lower.7ex\hbox{E}\kern-.125emX}}
    
\addto\captionsenglish{\renewcommand{\figurename}{Fig.}}

\addto\extrasenglish{%
}
    
\usepackage[square,numbers]{natbib}
\begin{document}

\title{Power System Architecture and Control for Green Hydrogen Production via Power Converter-less Photovoltaic-Electrolyser Integration}

\author{\IEEEauthorblockN{Aymeric Fabre}
\IEEEauthorblockA{\textit{Department of Electrical and Electronic Engineering} \\
\textit{The University of Melbourne}\\
Australia \\
aymeric.fabre@student.unimelb.edu.au}
\and
\IEEEauthorblockN{Glen Farivar}
\IEEEauthorblockA{\textit{Department of Electrical and Electronic Engineering} \\
\textit{The University of Melbourne}\\
Australia \\
glen.farivar@unimelb.edu.au}
\and
\IEEEauthorblockN{Andre Chambers}
\IEEEauthorblockA{\textit{Department of Mechanical Engineering} \\
\textit{The University of Melbourne}\\
Australia \\
andre.chambers@unimelb.edu.au}
}

\maketitle

\begin{abstract}
This paper proposes a power system architecture and control for efficient and low-cost green hydrogen production. The proposed system integrates photovoltaic (PV) sources directly with an electrolyser stack, thereby eliminating the need for traditional power converters. With the removal of traditional power converters, maximum power point tracking is achieved through dynamic switching of electrolyser cells in the stack, enabling load variation to maintain optimal voltage for maximum power output. The demonstration methodology involves comprehensive MATLAB Simulink analysis of the integrated system performance through controlled PV-electrolyser interactions.

\end{abstract}


\section{Introduction}
 
\subsection{Photovoltaic Arrays \& Maximum Power Point Tracking}

For years, photovoltaic (PV) arrays have been widely utilised in the renewable energy sector to capture sunlight and convert it into usable energy, such as for generating hydrogen. Depending on various factors, a PV array has an optimal output point known as the maximum power point (MPP). To precisely find this point, several MPP tracking (MPPT) algorithms have been devised \cite{A-New-Approach-for-Solar-Module-Temperature-Estimation-Using-the-Simple-Diode-Model}. One widely used algorithm is the perturb and observe method, which involves making small adjustments to the array's operating point to locate the MPP \cite{Comprehensive-Review-of-Conventional-and-Emerging-Maximum-Power-Point-Tracking-Algorithms-for-Uniformly-and-Partially-Shaded-Solar-Photovoltaic-Systems}. Another approach, known as the hill climbing method, employs a different strategy to achieve the same goal, by iteratively adjusting the operating point to move towards the steepest point of the power-voltage curve. These MPPT techniques are instrumental in maximising the energy output from PV arrays \cite{Comprehensive-Studies-on-Operational-Principles-for-Maximum-Power-Point-Tracking-in-Photovoltaic-Systems}. Traditionally, MPPT is achieved by regulating the PV array voltage irrespective of the load voltage. This necessity drives the utilisation of power converters. However, a power converter ultimately leads to significant energy losses and increased costs.

\subsection{Electrolysers in Renewable Energy Systems}

Electrolysers, a technology with an extensive history, have started to emerge in modern energy systems. Their primary function involves the separation of hydrogen and oxygen from water molecules. The extracted hydrogen can then be used as a green fuel \cite{Modeling-and-Simulation-of-Hydrogen-Energy-Storage-System-for-Power-to-gas-and-Gas-to-power-Systems}. Large renewable energy projects have integrated electrolysers, with the aim of extracting hydrogen to drive future power systems, such as in industries and hybrid transportation \cite{Grid-Forming-Services-From-Hydrogen-Electrolyzers}, \cite{Analytical-Modeling-and-Control-of-Grid-Scale-Alkaline-Electrolyzer-Plant-for-Frequency-Support-in-Wind-Dominated-Electricity-Hydrogen-Systems}, \cite{Virtual-Inertia-Response-and-Frequency-Control-Ancillary-Services-From-Hydrogen-Electrolyzers}, \cite{Optimal-design-of-hybrid-wind/photovoltaic-electrolyzer-for-maximum-hydrogen-production-using-imperialist-competitive-algorithm}. Electrolysers are often coupled with renewables to store excess renewable power when renewable energy production is abundant. The hydrogen can then be used during periods of low renewable generation, to meet network demands. Typically, electrolysers function within a voltage range of 1.5-2 V \cite{Modeling-and-Control-of-a-Renewable-Hybrid-Energy-System-With-Hydrogen-Storage}, \cite{Dynamic-Modeling-of-a-Pressurized-Alkaline-Water-Electrolyzer:-A-Multiphysics-Approach}. To achieve higher voltage and power levels, electrolysers are arranged in a stack \cite{A-comprehensive-review-of-alkaline-water-electrolysis-mathematical-modeling}, \cite{Dynamic-Electrical-Circuit-Modeling-of-a-Proton-Exchange-Membrane-Electrolyzer-for-Frequency-Stability-Resiliency-and-Sensitivity-Analysis-in-a-Power-Grid}. This work demonstrates that through the utilisation of electronic switching to add or remove electrolysers in the stack, the overall voltage can be fine-tuned to match the MPP voltage of a PV array, with no requirement of power converters. 

\subsection{Power Converter-less System Architecture}

Power converters are widely used in PV-Electrolyser systems, to separate the PV array voltage from the electrolyser stack voltage. The necessity of converters arises because the PV array operates optimally at its MPP voltage, which is not equal to the load voltage. However, power converters are costly, bulky, and represent a substantial inefficiency in the system through inherent resistive losses, switching losses, transformer losses etc... \cite{Parasitic-inductance-effect-on-switching-losses-for-a-high-frequency-Dc-Dc-converter}. The prevailing literature on this subject has explored alternatives to mitigate these losses, often suggesting the elimination of converters at the PV array stage. However, in many cases, converters are still prevalent at the electrolyser stage \cite{An-Innovative-Converterless-Solar-PV-Control-Strategy-for-a-Grid-Connected-Hybrid-PV/Wind/Fuel-Cell-System-Coupled-With-Battery-Energy-Storage}.  This paper introduces a methodology to seamlessly couple PV arrays with the electrolysis process, without the use of a power converter. Accordingly, the direct supply of power enables greater energy utilisation and minimised losses.
This paper is organised as follows: \autoref{sec:Methodology} introduces the proposed  power converter-less system architecture. \autoref{sec:Simulation Results} provides simulation results and cost analysis. Finally, conclusions are provided in \autoref{sec:Conclusion}.  

\section{Methodology}
\label{sec:Methodology}


\begin{figure}[t]
\centering
\resizebox{0.9\columnwidth}{!}{%
\begin{circuitikz}
\tikzstyle{every node}=[font=\normalsize]
\draw (6.25,9.5) to[C,l={ \normalsize $C_{e}$}] (8.75,9.5);
\draw (6.25,8) to[R,l={ \normalsize $R_{e}$}] (8.75,8);
\draw [](6.25,9.5) to[short] (6.25,8);
\draw [](8.75,9.5) to[short] (8.75,8);
\draw (4.5,8.75) to[D,l={ \normalsize $V_{e}$}] (6.25,8.75);
\draw [](4.5,10.5) to[short, -o] (5.5,10.5);
\draw [](9.5,10.5) to[short, -o] (6.25,10.5);
\draw [](4.5,8.75) to[short] (4.5,10.5);
\draw[] (4.5,8.75) to[short] (3.75,8.75);
\draw [](8.75,8.75) to[short] (10,8.75);
\draw [](9.5,10.5) to[short] (9.5,8.75);
\node [font=\Large] at (4,8.5) {$+$};
\node [font=\Large] at (9.75,8.5) {$-$};
\draw [short] (5.5,10.5) .. controls (5.75,10.75) and (5.75,10.75) .. (6,11);
\end{circuitikz}
}%
\caption{Electrolyser model.}
\label{fig:Electrolyser}
\end{figure}
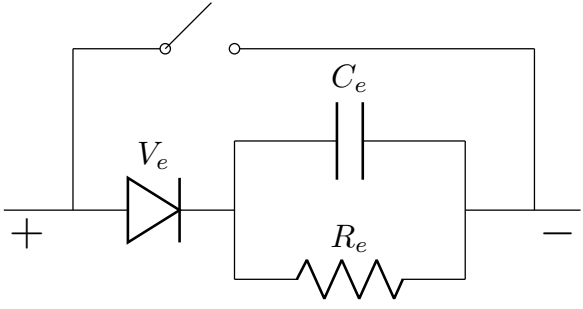

\begin{table}[t]
\centering
\caption{Electrolyser Parameter Values}
\label{tab:Electrolyser-Parameter-Values}
\resizebox{0.55\columnwidth}{!}{%
\begin{tabular}{ccc}
\toprule
\textbf{Parameter} & \textbf{Value} & \textbf{Unit} \\
\midrule
$V_e$              & 1.5            & V             \\
$C_e$              & 100            & mF            \\
$R_e$              & 700            & m$\Omega$    
\\
\bottomrule
\end{tabular}%
}
\end{table}

\begin{algorithm}[t]
\caption{MPPT Algorithm}\label{alg1}
\begin{algorithmic}[1]
\Require $n_{Total} = 30,~dn\in{\{-1,1\}},~G\in{\{0,1\}},~T\in{\mathbb{N}}$
\LeftComment { \textit{dn: electrolyser in stack, $(\mathrm{add,~remove}) = (1,-1)$)}}
\LeftComment { \textit{G: electrolyser state, $(\mathrm{on,~off}) = (0,1)$}}
\LeftComment { \textit{T: time electrolyser has spent in current state}}
\Procedure{UpdateStack}{$P,~G,~T,~dn,~n_{Active}$}
\State $T \gets T+1$
\If{$P_{curr} < P_{prev}$}
\LeftComment { \textit{Change direction of power curve hill climb}}
    \State $dn \gets -dn$
\EndIf

\If{$dn > 0$}
\LeftComment { \textit{Activate electrolyser that was inactive the longest}}
    \State $G(\max~T~\mathrm{for}~G=1) = 0$
    \State $T(\max~T~\mathrm{for}~G=1) = 0$
\ElsIf{$dn < 0$}
\LeftComment { \textit{Deactivate electrolyser that was active the longest}}
    \State $G(\max~T~\mathrm{for}~G=0) = 1$
    \State $T(\max~T~\mathrm{for}~G=0) = 0$
\EndIf

\State $n_{Active} \gets n_{Total} - \mathrm{sum}(G)$
\EndProcedure
\end{algorithmic}
\end{algorithm}

\begin{figure}[t]
\makebox[\columnwidth][c]{\includegraphics[width=1.1\columnwidth]{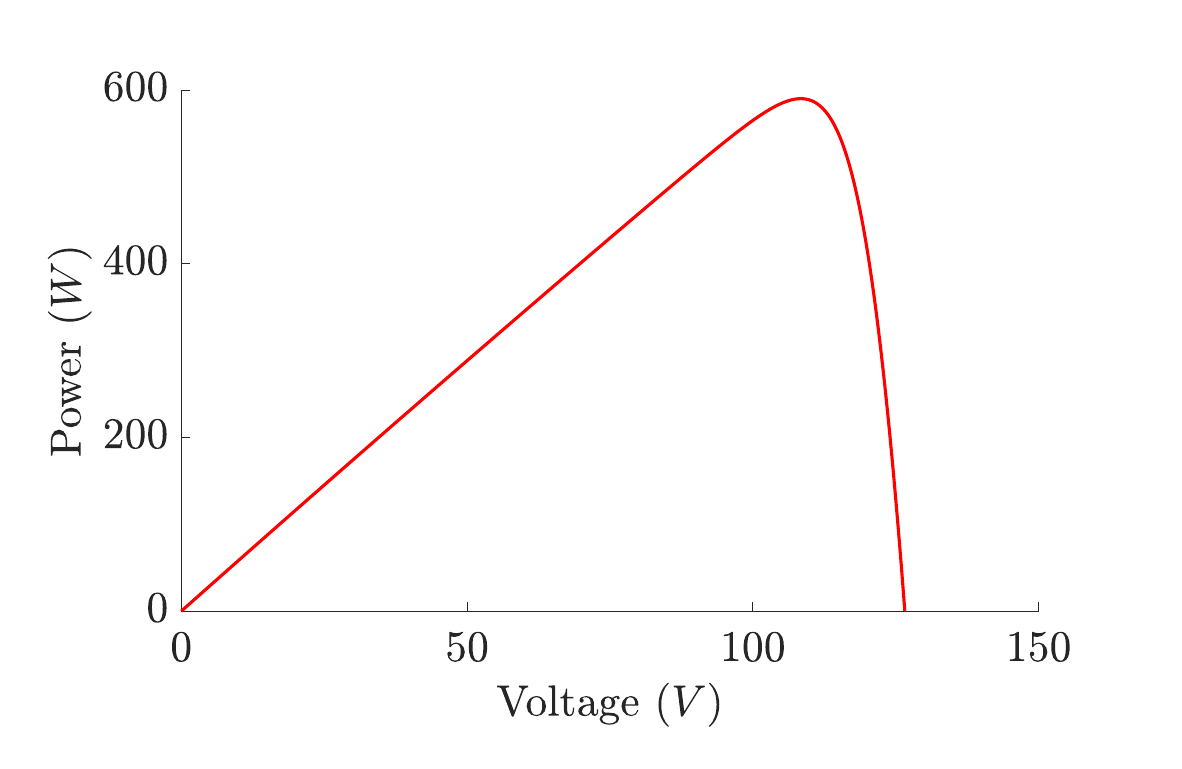}}
\vspace{-20pt}
\caption{PV Array Power vs Voltage (1000 $W\cdot{}m^{-2}$, 25 \textdegree C).}
\label{fig:Power_vs_Voltage}
\end{figure}

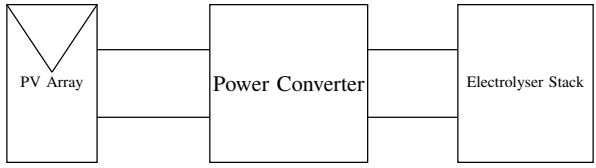
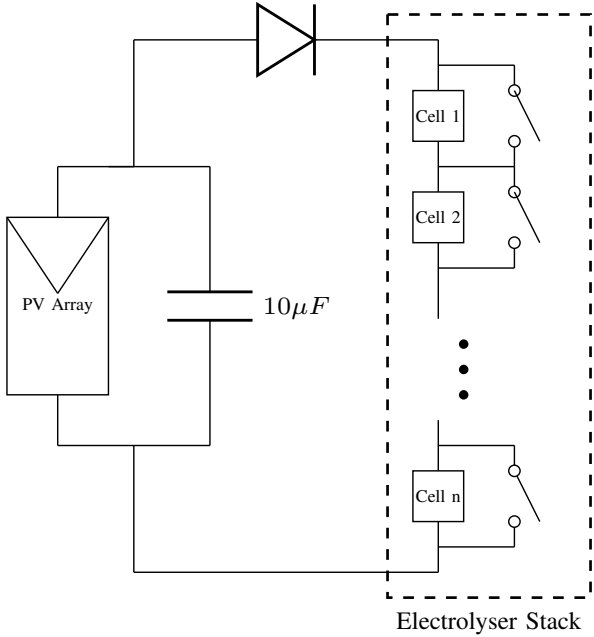
\begin{figure}[t]
    \begin{subfigure}[b]{1\columnwidth}
    \centering
\resizebox{0.9\columnwidth}{!}{%
\begin{circuitikz}
\tikzstyle{every node}=[font=\tiny]
\draw  (3.75,11.5) rectangle  node {\tiny PV Array} (4.75,9.75);
\draw [short] (3.75,11.5) to[short] (4.25,10.75);
\draw [short] (4.75,11.5) to[short] (4.25,10.75);
\draw [](4.75,11) to[short] (6,11);
\draw [](4.75,10.25) to[short] (6,10.25);
\draw  (6,11.5) rectangle  node {\scriptsize Power Converter} (7.75,9.75);
\draw  (8.75,11.5) rectangle  node {\tiny Electrolyser Stack} (10.25,9.75);
\draw [](7.75,11) to[short] (8.75,11);
\draw [](7.75,10.25) to[short] (8.75,10.25);
\end{circuitikz}
}%
\caption{Traditional power system architecture.}
\label{fig:Traditional diagram}
    \end{subfigure}
    \\\\
    \begin{subfigure}[b]{1\columnwidth}
    \centering
\resizebox{0.9\columnwidth}{!}{%
\begin{circuitikz}
\tikzstyle{every node}=[font=\scriptsize]
\draw  (6.25,10.25) rectangle  node {\tiny PV Array} (7.25,8.5);
\draw [, line width = 0.75pt , dashed] (10,12.25) rectangle  (12,6.5);
\draw [] (6.25,10.25) to[short] (6.75,9.5);
\draw [short] (7.25,10.25) to[short] (6.75,9.5);
\draw (7.5,12) to[D] (10.5,12);
\draw (8.25,10.75) to[C,l={ \scriptsize $10 \mu{}F$}] (8.25,8);
\draw [](6.75,10.25) to[short] (6.75,10.75);
\draw [](6.75,8.5) to[short] (6.75,8);
\draw [](7.5,6.75) to[short] (10.5,6.75);
\draw  (10.25,11.5) rectangle  node {\tiny Cell 1} (10.75,11);
\draw  (10.25,10.5) rectangle  node {\tiny Cell 2} (10.75,10);
\draw [](10.5,11.5) to[short] (10.5,12);
\draw [](10.5,11) to[short] (10.5,10.5);
\draw [](10.5,10.75) to[short] (11.25,10.75);
\draw [](10.5,11.75) to[short] (11.25,11.75);
\draw [](10.5,10) to[short] (10.5,9.25);
\draw  (10.25,7.75) rectangle  node {\tiny Cell n} (10.75,7.25);
\draw [](10.5,6.75) to[short] (10.5,7.25);
\node [font=\Huge] at (10.75,8.75) {.};
\node [font=\Huge] at (10.75,9) {.};
\node [font=\Huge] at (10.75,8.5) {.};
\draw [](10.5,9.75) to[short] (11.25,9.75);
\draw [](10.5,7) to[short] (11.25,7);
\draw [](10.5,8.25) to[short] (10.5,7.75);
\draw [](10.5,8) to[short] (11.25,8);
\draw [](6.75,10.75) to[short] (8.25,10.75);
\draw [](6.75,8) to[short] (8.25,8);
\draw [](7.25,10.75) to[short] (7.5,10.75);
\draw [](7.5,10.75) to[short] (7.5,12);
\draw [](7.5,8) to[short] (7.5,6.75);
\draw [](11.25,11.5) to[short, o-] (11.25,11.75);
\draw [](11.25,10.75) to[short, -o] (11.25,11);
\draw [short] (11.25,11.5) to[short] (11.5,11);
\draw [](11.25,10.5) to[short, o-] (11.25,10.75);
\draw [](11.25,9.75) to[short, -o] (11.25,10);
\draw [short] (11.25,10.5) to[short] (11.5,10);
\draw [short] (11.25,7.75) to[short] (11.5,7.25);
\draw [](11.25,7.75) to[short, o-] (11.25,8);
\draw [](11.25,7) to[short, -o] (11.25,7.25);
\node [font=\scriptsize] at (11,6.25) {Electrolyser Stack};
\end{circuitikz}
}%
\caption{Power converter-less system architecture.}
\label{fig:System Diagram}
    \end{subfigure}
    \caption{System diagrams.}
    \label{fig:System diagrams}
\end{figure}

In this study, proton exchange membrane (PEM) electrolysers were modelled as a capacitor and resistor in parallel, with a diode in series as can be seen in \autoref{fig:Electrolyser} \cite{Proton-Exchange-Membrane-Electrolyzer-Emulator-for-Power-Electronics-Testing-Applications}. A PV array is conveniently modelled using a single diode circuit through the embedded Simulink PV array. The PV MPP varies with the temperature and irradiance, therefore, PV systems often incorporate a MPPT algorithm to maximise the power output. \autoref{fig:Power_vs_Voltage} displays the power-voltage characteristic of the PV array employed in \autoref{sec:Simulation Results}.

In conventional system, a DC-DC power converter decouples the dynamics of the PV array and electrolyser stack. The application of a typical perturb and observe MPPT algorithm involves making small adjustments to the PV-side voltage to locate the MPP, regardless of the electrolyser-side voltage. In contrast, the proposed system illustrated in \autoref{fig:System Diagram} eliminates the need for a power converter and instead features switchable electrolyser cells. The system contains a diode to prevent reverse current through the PV array and n-cell electrolyser stack, in parallel with a 10 $\mu$F smoothing capacitor to reduce abrupt voltage changes. Without a power converter, this system achieves MPPT by selectively switching ON/OFF the electrolyser cells, a process detailed in the rest of this section.

To ensure an equal distribution of workload among the electrolysers, a novel MPPT algorithm was designed. \autoref{alg1} selectively activates electrolysers that have remained inactive for the longest duration and deactivates those that have been operational for the longest time. This approach guarantees a balanced utilisation of the electrolysers, such that they each contribute equally over time, in addition to balancing the lifespan of all electrolyser cells. The MPPT controller was triggered to update the electrolyser stack every second. In future works, the model can be tuned to more accurately represent the amount of time an electrolyser requires to begin producing hydrogen.

\vspace{40pt}





\section{Simulation Results}
\label{sec:Simulation Results}

The proposed power system architecture was implemented in MATLAB Simulink. The Simulink integrated PV array model was used for simplicity. The array had a short circuit current of 5.83 A and a maximum power point of 108.4 V, 590 W at 1000 $W\cdot{}m^{-2}$ and 25 \textdegree C. The parameters of the electrolyser model are provided in \autoref{tab:Electrolyser-Parameter-Values}. To demonstrate functionality, a stack of 30 electrolysers was employed. When all 30 electrolysers were operational, the system could draw 120 V allowing the MPPT algorithm to converge to the designated maximum power point, without necessitating the activation of all electrolysers simultaneously. Two simulations were completed to demonstrate the nominal operation and the system's response to a change in irradiance. Lastly, a high-level cost analysis is demonstrated.

\subsection{Nominal Operation Startup Test}
\label{sec:Nominal Operation}

\begin{figure*}[t!]
    \begin{subfigure}[b]{1\columnwidth}
        \includegraphics[width=1\columnwidth]{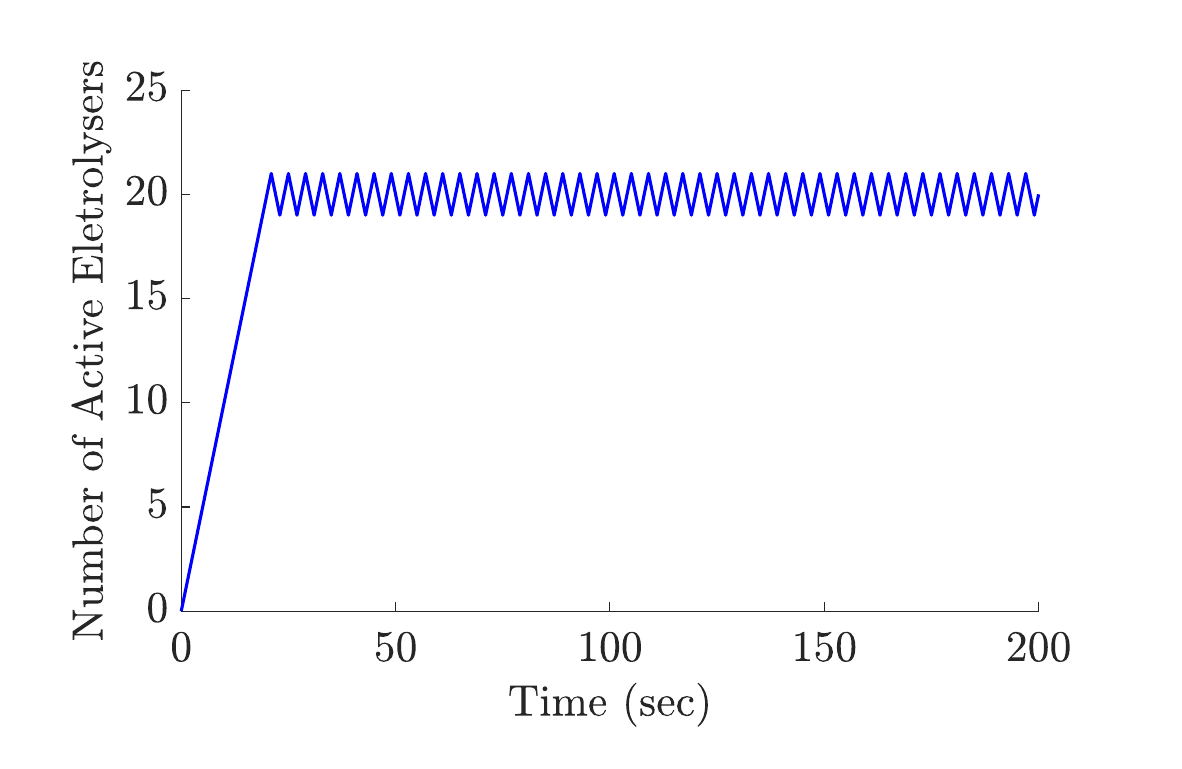}
        \caption{Number of active electrolysers under startup test.}
        \label{fig:1000_number_of_active_electrolysers_vs_time}
    \end{subfigure}
    \hfill
    \begin{subfigure}[b]{1\columnwidth}
        \includegraphics[width=1\columnwidth]{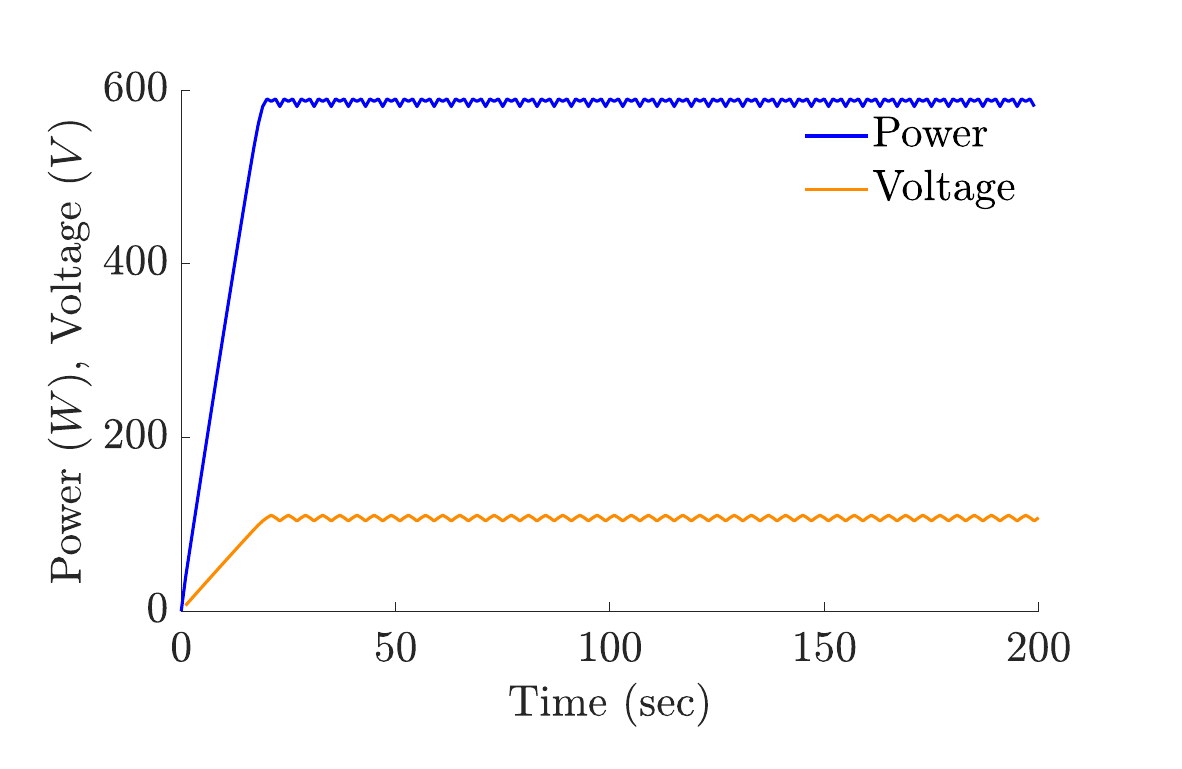}
        \caption{Power and voltage under startup test.}
        \label{fig:1000_power_and_voltage_vs_time}
    \end{subfigure}

    \begin{subfigure}[b]{1\columnwidth}
        \includegraphics[width=1\columnwidth]{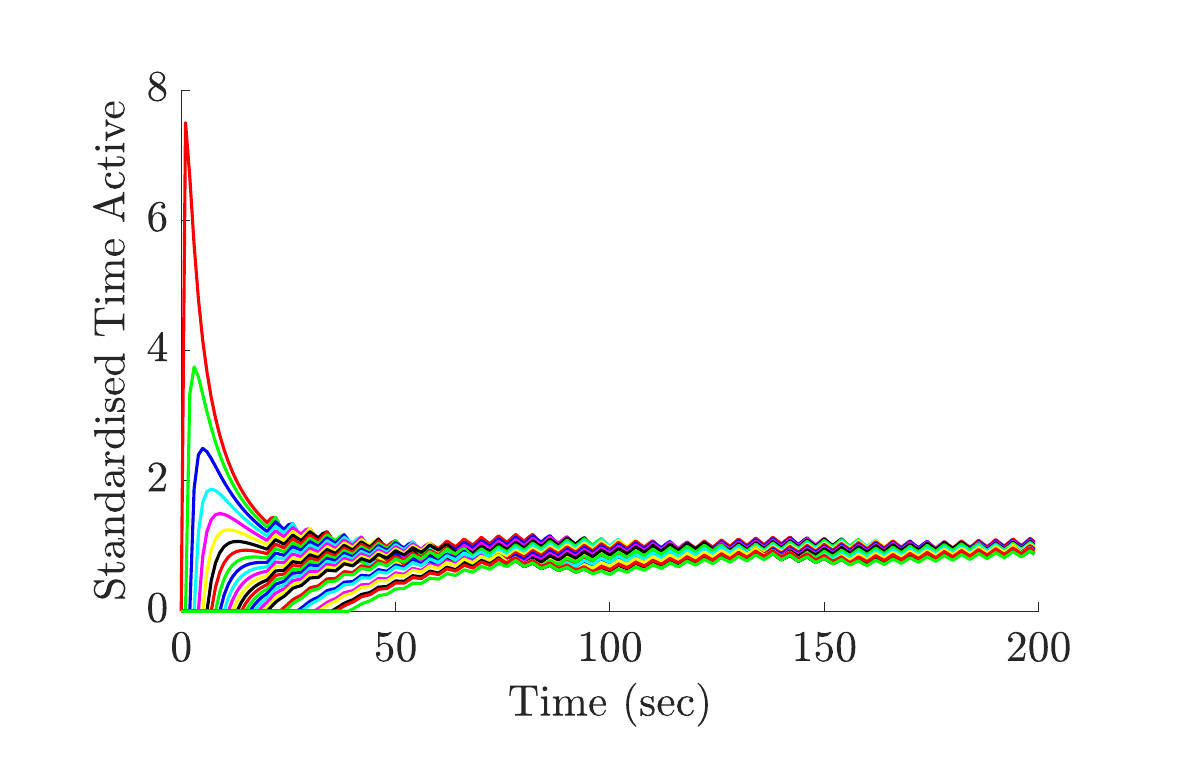}
        \caption{Standardised time active for all electrolysers under startup test.}
        \label{fig:1000_standardised_time_active_vs_time}
    \end{subfigure}
    \hfill
    \begin{subfigure}[b]{1\columnwidth}
        \includegraphics[width=1\columnwidth]{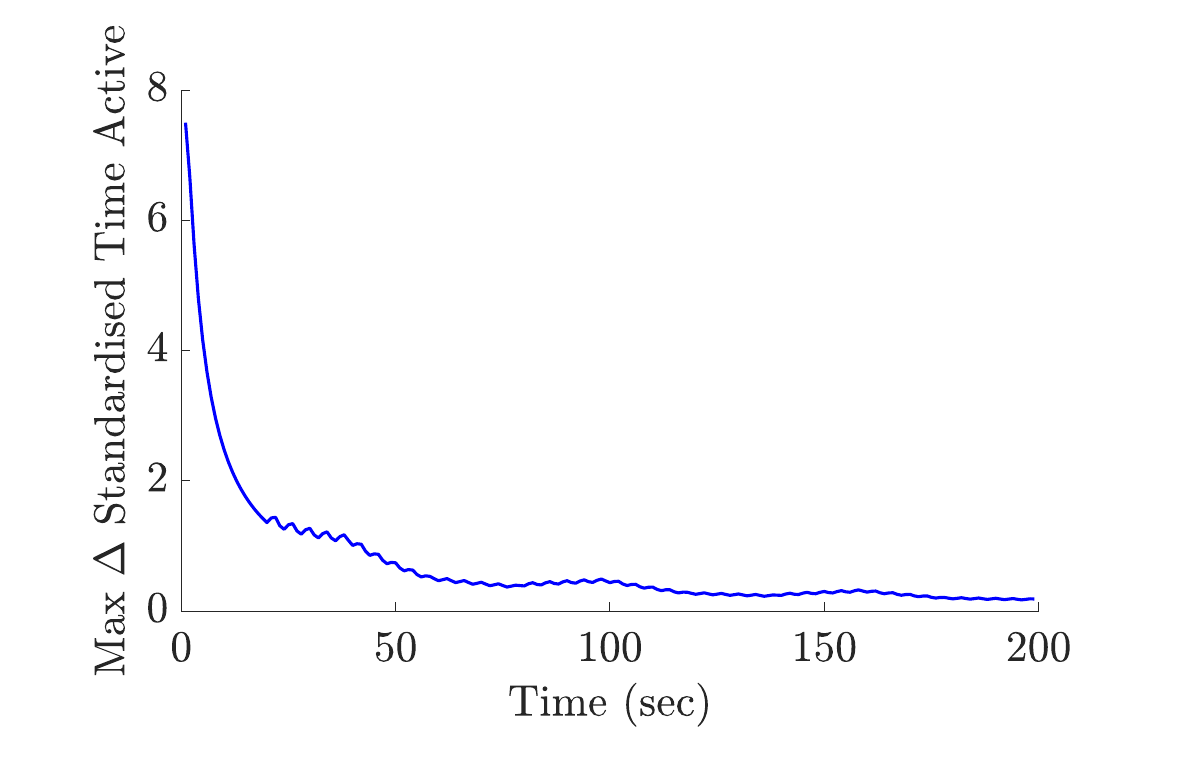}
        \caption{Max $\Delta$ standardised time active under startup test.}
        \label{fig:1000_max_delta_standardised_time_active_vs_time}
    \end{subfigure}
    \caption{Startup test simulation results.}
    \label{fig:Startup test simulation results}
\end{figure*}

The first simulation demonstrates the nominal operation of the system with an irradiance set to 1000 $W\cdot{}m^{-2}$ and the temperature set to 25 \textdegree C. \autoref{fig:1000_number_of_active_electrolysers_vs_time} displays the nominal system startup process, whereby electrolysers are switched on progressively until 20 are activated. At this point, the power and voltage, shown in \autoref{fig:1000_power_and_voltage_vs_time}, reached their respective MPP values. Hence, the number of electrolysers active oscillate between 19 to 21, as the algorithm continues to track the MPP.

To demonstrate the effectiveness of the proposed MPPT algorithm, the standardised time spent ON for each electrolyser was found using the following equations:
\\\\
Proportion of time an electrolyser has been active:
\\\\
$\mathrm{TA~(Time~Active~\%)} = \frac{\mathrm{Time~Spent~ON}}{\mathrm{Current~Simulation~Time}}$
\\\\
Standardised proportion of time an electrolyser has been active, such that it converges to 1:
\\\\
$\mathrm{STA~(Standardised~Time~Active)} = \mathrm{TA} \div{} \frac{n_{Active}}{n_{Total}}$
\\\\
The maximum delta in time spent on between the least and most active electrolysers is expected to converge to 0:
\\\\
Max $\Delta{}$ STA $= \max(\mathrm{STA}) - \min(\mathrm{STA})$

The effect of this standardisation is illustrated in \autoref{fig:1000_standardised_time_active_vs_time}, where the standardised time active converges to 1 for all electrolysers. The asymptotic behaviour to 1 is significant because it empirically proves the MPPT algorithm equally distributes the workload across all electrolysers over time. \autoref{fig:1000_max_delta_standardised_time_active_vs_time} demonstrates the MPPT algorithm correctly switching on and off electrolysers to ensure they are all active for a similar amount of time, as the difference between the electrolysers which have been active for the longest and shortest duration, converges to 0.


\subsection{Response to Step Change in Irradiance}

\begin{figure*}[t]
    \begin{subfigure}[b]{1\columnwidth}
        \includegraphics[width=1\columnwidth]{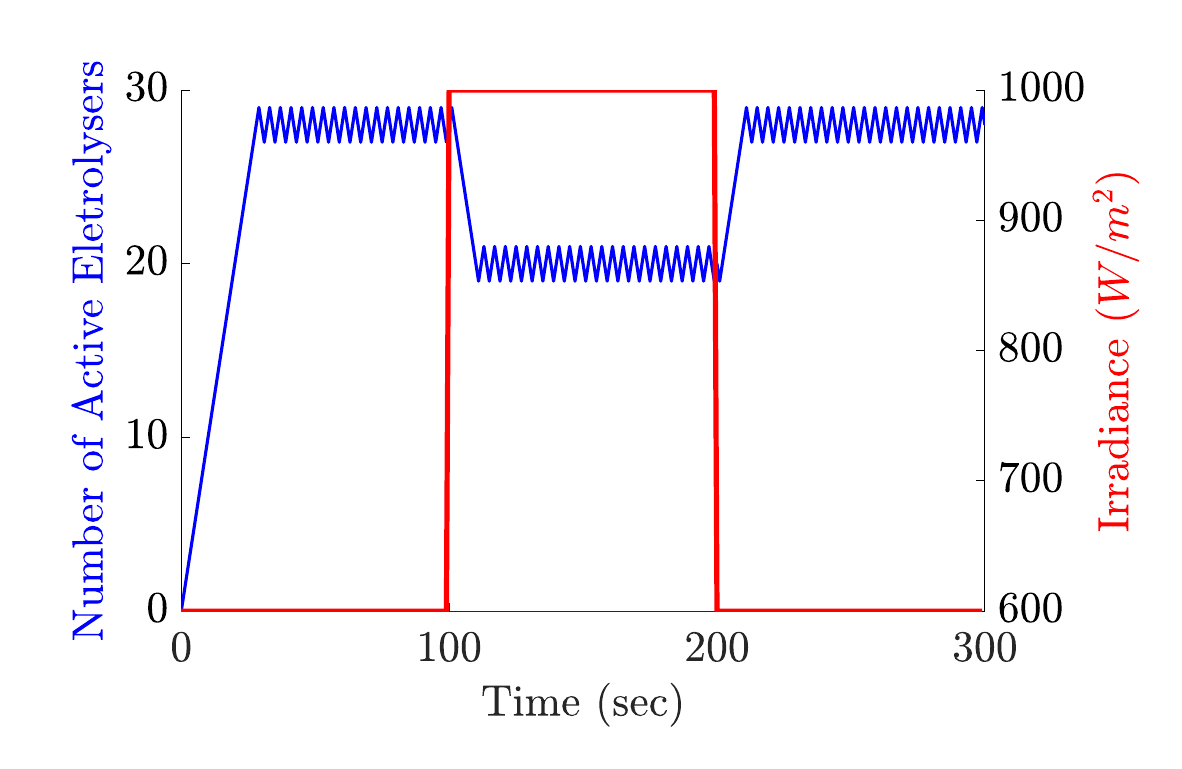}
        \caption{Number of active electrolysers under $\Delta$Irradiance.}
        \label{fig:600-1000-600_number_of_active_electrolysers_vs_time}
    \end{subfigure}
    \hfill
    \begin{subfigure}[b]{1\columnwidth}
        \includegraphics[width=1\columnwidth]{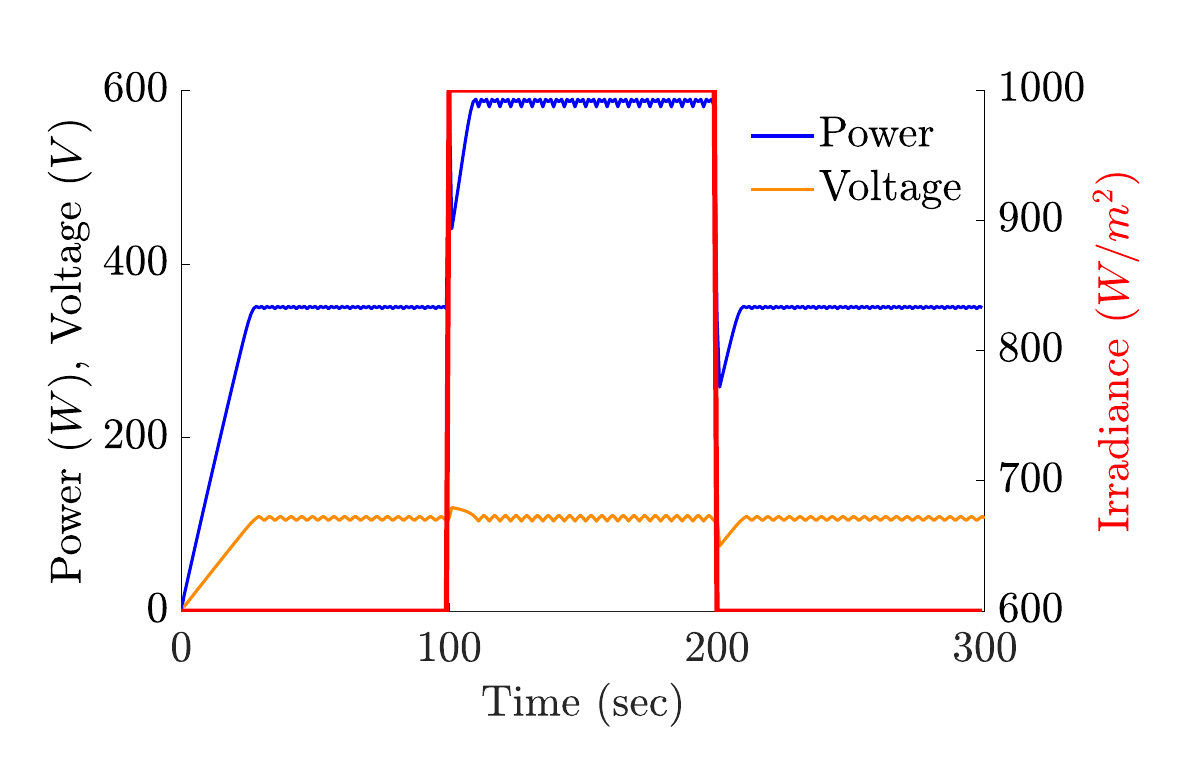}
        \caption{Power and voltage under $\Delta$Irradiance.}
        \label{fig:600-1000-600_power_and_voltage_vs_time}
    \end{subfigure}

    \begin{subfigure}[b]{1\columnwidth}
        \includegraphics[width=1\columnwidth]{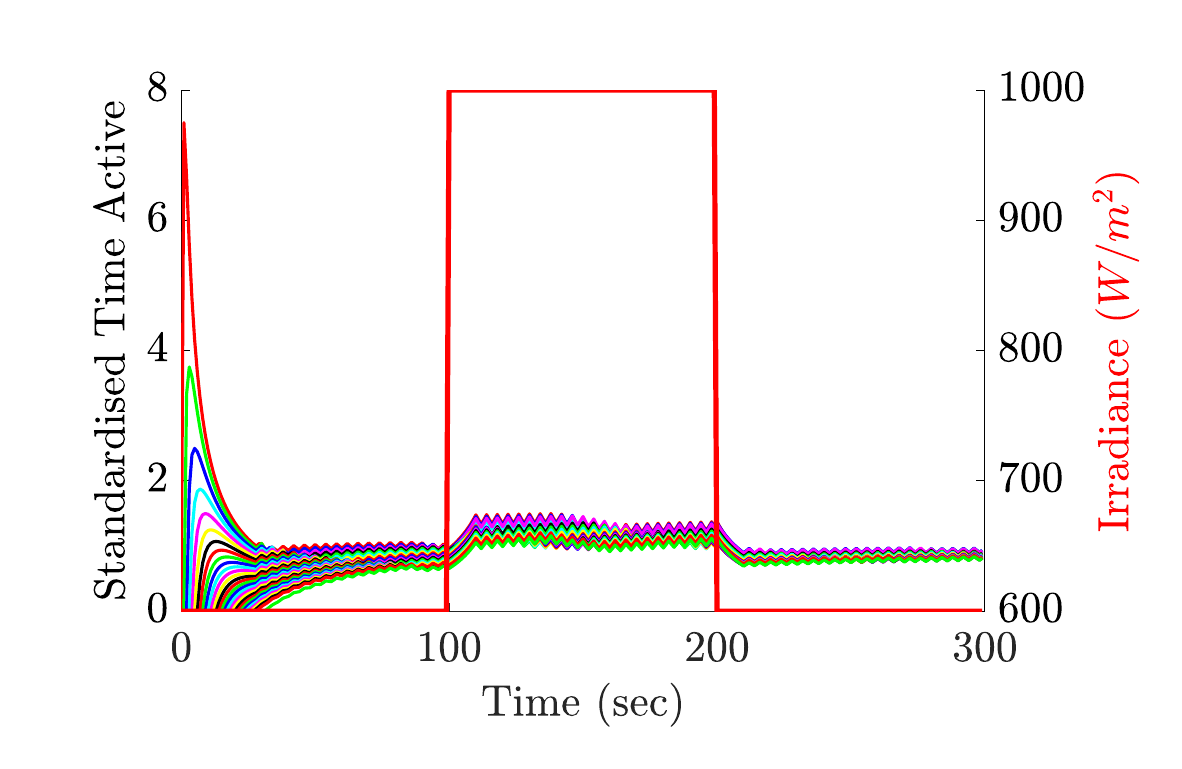}
        \caption{Standardised time active for all electrolysers under $\Delta$Irradiance.}
        \label{fig:600-1000-600_standardised_time_active_vs_time}
    \end{subfigure}
    \hfill
    \begin{subfigure}[b]{1\columnwidth}
        \includegraphics[width=1\columnwidth]{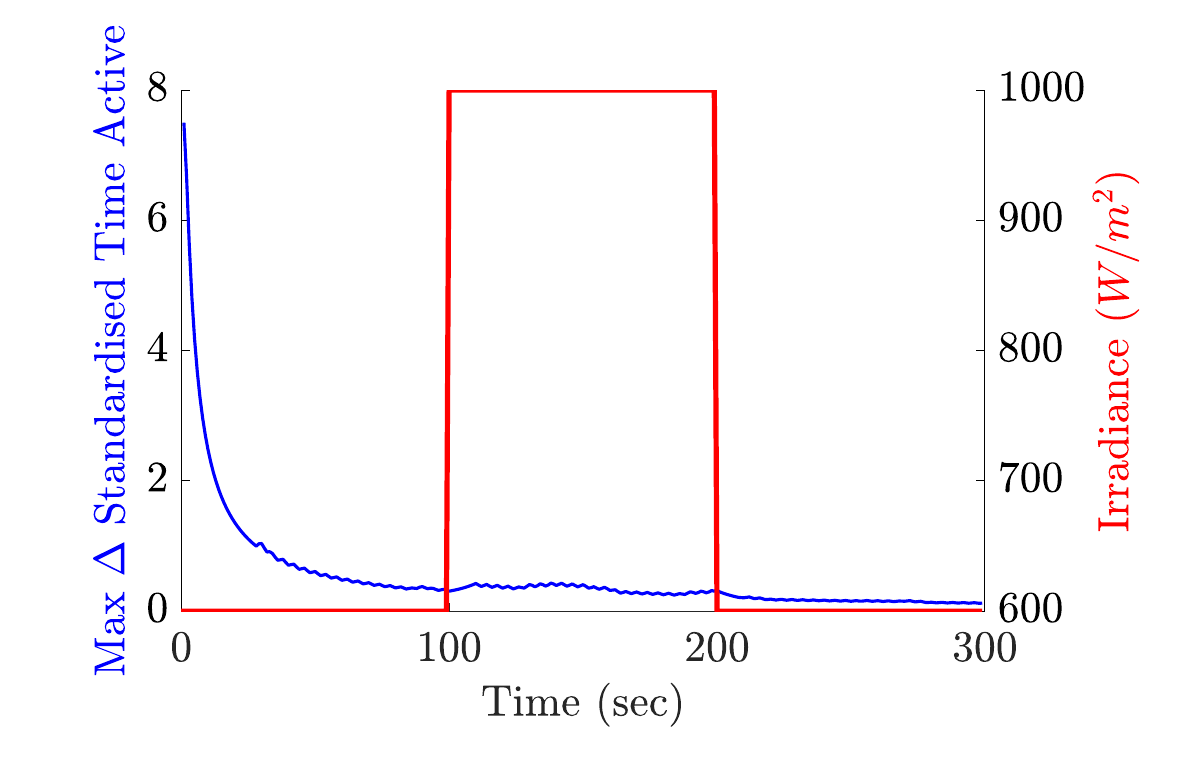}
        \caption{Max $\Delta$ standardised time active under $\Delta$Irradiance.}
        \label{fig:600-1000-600 max delta standardised time active vs time}
    \end{subfigure}
    \caption{Irradiance step change simulation results.}
    \label{fig:Irradiance step change simulation results}
\end{figure*}





In this section, a simulated change in irradiance is conducted to demonstrate the system response. The temperature was set to 25 \textdegree C, and the irradiance was set as follows:
\\\\
$Irradiance(t) = \begin{cases} 
    600 & 0\leq t< 100 \\
    1000 & 100\leq t\leq 200 \\
    600 & 200< t\leq 300 
\end{cases}, ~~~ (W\cdot{}m^{-2})$
\\\\\\
\autoref{fig:600-1000-600_number_of_active_electrolysers_vs_time} displays the nominal system startup, however, a noteworthy observation is that the system now requires 28 electrolysers active to reach the MPP, since the irradiance is considerably lower than in \autoref{sec:Nominal Operation}. At MPP, the number of electrolysers active oscillates between 27 to 29, and the algorithm continues to track the maximum power point. The power and voltage, shown in \autoref{fig:600-1000-600_power_and_voltage_vs_time}, are able to swiftly reach their respective MPP values, irrespective of the change in irradiance.


The standardisation of the electrolyser's time active in \autoref{fig:600-1000-600_standardised_time_active_vs_time} reveals the change in irradiance has negligible effect on the balancing of electrolyser usage. This observation is further supported by \autoref{fig:600-1000-600 max delta standardised time active vs time}, which demonstrates a smooth convergence to 0 for the difference between the most and least used electrolysers.

\subsection{Cost Efficiency Analysis}

In regard to cost efficiency, a simple analysis can be done for the average cost of power. The proposed scenario involves a 10 kW PV array, \$2,000 upfront cost for a standard power converter, a system lifespan of 10 years, and the cost of electricity with a power converter: \$0.15 /kWh. Over 10 years, the total cost of energy production with the converter would be:
\\\\
$\mathrm{Cost~with~Converter} = \$2,000~(\mathrm{upfront~cost}) + 10 \times (10,000~\mathrm{kWh} \times \$0.15~\mathrm{per~kWh}) = \$17,000$
\\\\
In the power converter-less system, capable of maintaining MPPT, the initial upfront converter cost would be eliminated, along with its associated power losses. Hence a realistic and feasible cost savings of 1 cent per kWh could be observed, assuming the converter is $\sim$90\% efficient, yielding the following cost:
\\\\
$\mathrm{Cost~without~Converter} = 10 \times (10,000~\mathrm{kWh} \times \$0.14~\mathrm{per~kWh}) = \$14,000$
\\\\
$\mathrm{Cost~Savings} = \$17,000-\$14,000 = \$3,000$
\\\\
$\mathrm{Percentage~Cost~Savings} = \frac{\$3,000}{\$17,000} \times 100\% \approx \boxed{18\%}$
\\\\
Hence a potential 18\% decrease in cost could be achieved from using this converter-less system. However, it is important to acknowledge the real-world implementation of this system would require more cells in a stack and a few additions to the electrolyser stack in the form of electronically controlled relay switches for each electrolyser cell.

\section{Conclusion}
\label{sec:Conclusion}

Overall, the proposed power system architecture and control strategy for PV-electrolyser systems has successfully eliminated the requirement for power converters and consequently has significantly reduced power losses and costs. This approach has proven its capability of achieving MPP in conjunction with efficiently balancing the utilisation of individual electrolyser cells within the stack. The implications of these novel findings pave the way for future renewable energy projects and contribute to drive down the cost of the global energy transition to clean energy sources.
\\\\
A potential drawback of the system architecture's is that it is not easily scalable. The simulation model outlined in this paper operates at a voltage of approximately 100 V, which is far less than large scale renewable energy solar farms, often operating in the kilo-volt range. To keep the current power system architecture and still manage to reach such MPP voltages would require multiplying the number of electrolysers by a factor of 10. Although this may be challenging to achieve, there are numerous renewable energy projects which are planning to construct large amounts of electrolysers. One such example is the planned Western Green Energy Hub (WGEH) in Western Australia, proposing 50 GW of wind and solar capacity and producing 3.5 million tonnes of green hydrogen yearly. The proposed power converter-less system would excel in this type of project, whereby there would be an abundance of electrolysers.
\\\\
Ultimately, our study contributes to advancing green H2 production methods and exemplifies a notable step towards optimising renewable energy integration within power systems.

\section*{Acknowledgment}

The authors of this paper thank the researchers from the Melbourne Energy Institute for their academic support.

\bibliography{bibliography.bib}

\end{document}